\documentclass[conference,letterpaper]{IEEEtran}
\usepackage{amsfonts,amsbsy,bm,amsmath}
\usepackage[nolist]{acronym}
\usepackage{mathrsfs} 
\usepackage{graphicx,cite,amssymb,mathtools,amsthm}
\usepackage{stfloats}
\usepackage{xcolor}
\usepackage{tikz}
\usepackage{pgfplots}
\usepackage{pgf}
\mathtoolsset{showonlyrefs}
\usepackage{bbm}
\allowdisplaybreaks

\usepackage{subcaption}
\newcommand{\MCC}{}

\newcommand{\code}{\mathcal{C}}

\newcommand{\G}{\boldsymbol{G}}
\newcommand{\GK}{\boldsymbol{\mathsf{K}}}

\newtheorem{prop}{Proposition}

\newtheorem{example}{Example}

\definecolor{lightblue}{rgb}{0,.5,1}
\definecolor{normemph}{rgb}{0,.2,0.6}
\definecolor{supremph}{rgb}{0.6,.2,0.1}
\definecolor{lightpurple}{rgb}{.6,.4,1}
\definecolor{gold}{rgb}{.6,.5,0}
\definecolor{orange}{rgb}{1,0.4,0}
\definecolor{hotpink}{rgb}{1,0,0.5}
\definecolor{newcolor2}{rgb}{.5,.3,.5}
\definecolor{newcolor}{rgb}{0,.3,1}
\definecolor{newcolor3}{rgb}{1,0,.35}
\definecolor{darkgreen1}{rgb}{0, .35, 0}
\definecolor{darkgreen}{rgb}{0, .6, 0}
\definecolor{darkred}{rgb}{.75,0,0}
\definecolor{midgray}{rgb}{.8,0.8,0.8}
\definecolor{darkblue}{rgb}{0,.25,0.6}

\definecolor{lightred}{rgb}{1,0.9,0.9}
\definecolor{lightblue}{rgb}{0.9,0,0.0}
\definecolor{lightpurple}{rgb}{.6,.4,1}
\definecolor{gold}{rgb}{.6,.5,0}
\definecolor{orange}{rgb}{1,0.4,0}
\definecolor{hotpink}{rgb}{1,0,0.5}
\definecolor{darkgreen}{rgb}{0, .6, 0}
\definecolor{darkred}{rgb}{.75,0,0}
\definecolor{darkblue}{rgb}{0,0,0.6}

\definecolor{bgblue}{RGB}{245,243,253}
\definecolor{ttblue}{RGB}{91,194,224}

\definecolor{dark_red}{RGB}{150,0,0}
\definecolor{dark_green}{RGB}{0,150,0}
\definecolor{dark_blue}{RGB}{0,0,150}
\definecolor{dark_pink}{RGB}{80,120,90}

\begin{acronym}
	\acro{app}{a posteriori probability}
	\acro{AWGN}{input additive white Gaussian noise}
	\acro{biAWGN}{binary-input additive white Gaussian noise}
	\acro{B-DMC}{binary-input discrete memoryless channel}
	\acro{B-DMSC}{binary-input discrete memoryless symmetric channel}
	\acro{BMS}{binary memoryless symmetric}
	\acro{BCJR}{Bahl, Cocke, Jelinek, and Raviv}
	\acro{BEC}{binary erasure channel}
	\acro{BER}{bit error rate}
	\acro{BLEP}{block error probability}
	\acro{BP}{belief propagation}
	\acro{BSC}{binary symmetric channel}
	\acro{CER}{codeword error rate}
	\acro{CN}{check node}
	\acro{CRC}{cyclic redundancy check}
	\acro{DE}{density evolution}
	\acro{eBCH}{extended Bose–Chaudhuri–Hocquenghem}
	\acro{eHamming}{extended Hamming}
	\acro{GA}{Gaussian approximation}
	\acro{GRAND}{guessing random additive noise decoding}
	\acro{HDD}{hard-decision decoding}
	\acro{i.i.d.}{independent and identically distributed}
	\acro{IO-WE}{input-output weight enumerator}
	\acro{IR-WE}{input-redundancy weight enumerator}
	\acro{IO-WEF}{input-output weight enumerating function}
	\acro{IR-WEF}{input-redundancy weight enumerating function}
	\acro{JIO-WE}{joint IO-WE}
	\acro{JIR-WE}{joint IR-WE}
	\acro{JWE}{joint WE}
	\acro{LDPC}{low-density parity-check}
	\acro{LHS}{left-hand side}
	\acro{LLR}{log-likelihood ratio}
	\acro{MAP}{maximum a posteriori}
	\acro{MC}{metaconverse}
	\acro{ML}{maximum likelihood}
	\acro{PAC}{polarization-adjusted convolutional}
	\acro{PC}{product code}
	\acro{pdf}{probability density function}
	\acro{RCB}{random coding bound}
	\acro{RCUB}{random coding union bound}
	\acro{RM}{Reed--Muller}
	\acro{RM-polar}{Reed--Muller polar}
	\acro{RHS}{right-hand side}
	\acro{RV}{random variable}
	\acro{SPC}{single parity-check}
	\acro{SC}{successive cancellation}
	\acro{SCC}{super component codes}
	\acro{SCL}{successive cancellation list}
	\acro{SISO}{soft-input soft-output}
	\acro{SNR}{signal-to-noise ratio}
	\acro{UB}{union bound}
	\acro{TUB}{truncated union bound}
	\acro{VN}{variable node}
	\acro{WE}{weight enumerator}
	\acro{WEF}{weight enumerating function}
\end{acronym}

\begin{document}

	\title{Precoded Polar Product Codes}
%	\author{Mustafa Cemil Co\c{s}kun, \IEEEmembership{Student Member, IEEE}, Thomas Jerkovits, \IEEEmembership{Student Member, IEEE}, Gianluigi Liva, \IEEEmembership{Senior Member, IEEE}
%	\thanks{This work was supported in part by the research grant "Efficient Coding and Modulation for Satellite Links with Severe Delay Constraints" funded by Munich Aerospace e.V.}
%	\thanks{Mustafa Cemil Co\c{s}kun, Thomas Jerkovits and Gianluigi Liva are with the Institute of Communications and	Navigation of the German Aerospace Center (DLR), M\"unchner Strasse 20, 82234 We{\ss}ling, Germany (email: \{mustafa.coskun,gianluigi.liva,thomas.jerkovits\}@dlr.de).}
%	} 
	
		\author{\IEEEauthorblockN{Mustafa Cemil Co\c{s}kun}
		
		\IEEEauthorblockA{Nokia Bell Labs\\ Murray Hill, NJ, USA \\
			Email: mustafa.coskun@nokia-bell-labs.com
		}}

	\maketitle

	\begin{abstract}
		Precoded polar product codes are proposed, where selected component codes enable successive cancellation list decoding to generate bit-wise soft messages efficiently for iterative decoding while targeting optimized distance spectrum as opposed to eBCH or polar component codes. Rate compatibility is a byproduct of $1$-bit granularity in the component code design.
	\end{abstract}

	%\vspace{-5mm}
\section{Introduction}\label{sec:intro}

Product codes are introduced as the first deterministic codes with vanishing bit-error probability and positive rates\cite{Elias:errorfreecoding54}, where their component codes are decoded in a one-sweep fashion starting from the codes of first dimension up to those of the last one. Their iterative \ac{HDD} was discussed in \cite{Abramson}, but product codes got attention only after \ac{SISO} component code decoding is used \cite{Pyndiah98} similar to turbo codes\cite{Berrou93}.

Product code variants have distinctive features such as approaching capacity at high rates\cite{NHB97,Pyndiah98,Li04,JPN17}, suitability for fast parallel decoding architectures\cite{AM02} and having relatively large minimum distance (i.e., potentially low error floor)\cite{Chiaraluce}. They are, usually, constructed from short and powerful algebraic codes, which  enables the use of \ac{SISO}\cite{Pyndiah98} (e.g., \ac{BCJR}\cite{BCJR}) or algebraic (e.g., bounded-distance) \cite{Abramson,KoetterPC,Haeger2018} decoders for the component codes to address high data rate applications. For an overview of product code variants, we refer a curious reader to \cite{MAS16:TPC}.

Polar codes are the first class of codes achieving the capacity of \ac{BMS} channels under low-complexity \ac{SC} decoding\cite{arikan2009channel}. However, their initial performance was not competitive to state-of-art codes at short- to moderate-length regime. Later, their modifications, e.g., \ac{CRC}-aided polar codes\cite{tal15}, polar codes with dynamic frozen bits\cite{TM16} and \ac{PAC} codes\cite{arikan2019sequential}, shown to have excellent performance under \ac{SCL} decoding algorithm\cite{tal15} thanks to the improved distance spectrum coupled with improved decoding, leading to one of the most powerful code classes for the short blocklength regime\cite{Coskun18:Survey}. In the following, we will focus on polar codes with dynamic frozen bits and refer to them as precoded polar codes as appeared in \cite{MV20}. 

Recently, product codes with polar component codes, which we call polar product codes, are proposed\cite{KCW+2018,BCL19} to reduce the decoding latency compared to \ac{SC}-based decoding of an entire polar code. To this end, their iterative (or turbo) decoding is considered, where soft information is extracted via modification to \ac{SC}-based \ac{HDD} similar to \cite{Pyndiah98}. In addition, \cite{BCL19} exploited the fact that the resulting product code can be seen as a polar code by using \ac{SC}-based decoding of the entire product code if the iterative decoding does not converge to a valid codeword. Motivated by the considered applications, both works focused on very high rate polar product codes. The performance significantly degrades if one adopts low-rate polar component codes due to the poor distance properties even if near \ac{ML} decoding is possible. 

In this work, product codes with precoded polar component codes, namely precoded polar product codes, are proposed, where their distance spectrum is significantly improved compared to polar product codes. To be able to use SCL decoding efficiently for bit-wise soft output generation, component codes are chosen particularly suited for \ac{SCL} decoding. The potential improvement in the performance is analyzed via the distance spectrum restricted to the minimum distance terms, which is verified via simulations. Under iterative decoding, the proposed codes can perform the same as product codes with \ac{eBCH} component codes even when \ac{BCJR} decoding is used for \ac{eBCH} codes, which are competitive to 5G \ac{LDPC} codes for the considered cases. Note finally that precoded polar product codes are naturally rate adaptive thanks to the $1$-bit granularity of the considered component codes. 

The work is organized as follows. Preliminaries are given in Section \ref{sec:prelim}. Section \ref{sec:dpolar_product} describes precoded polar product codes together with distance as well as the decoding algorithm. After providing the numerical results in Section \ref{sec:numerical_results}, conclusions follow in Section \ref{sec:conc}.
	\section{Preliminaries}\label{sec:prelim}

\subsection{Notation and System Model}
In the following, $ x_a^b $ is used to denote vector $ (x_a, x_{a+1}, \dots, x_b) $ where $ b > a $. If $b<a$, then $x_{a}^{b}$ is void. We use $[N]$ for the set $\left\{ 1,2,\ldots N\right\}$. Capital bold letters, e.g., $\boldsymbol{X}$, are used for matrices, where the dimension becomes clear from the context. In addition, capital bold letters with sans-serif font are used for square transform matrices, to clearly distinguish them from other matrices, e.g., generator matrices. The Kronecker product of two matrices $\boldsymbol{X}$ and $\boldsymbol{Y}$ is
\begin{equation*}
	\boldsymbol{X}\otimes \boldsymbol{Y} \triangleq \begin{bmatrix}
		x_{1,1}\boldsymbol{Y} & x_{1,2}\boldsymbol{Y} & \dots \\ 
		x_{2,1}\boldsymbol{Y} & x_{2,2}\boldsymbol{Y} & \dots \\
		\vdots & \vdots &\ddots
	\end{bmatrix}. \label{eq:Kronecker}
\end{equation*}
%where $x_{i,j}$ is the element of $\boldsymbol{X}$ in the $i$th row and $j$th column.

%We use capital letters, e.g., $X$, for \acp{RV} and lower-case counterparts, e.g., $x$, for their realizations. For random vectors, similar notation above is used, e.g., we use $X_a^b$ to denote the random vector $(X_a, X_{a+1},\ldots X_{b})$.
We use capital letters, e.g., $X$, for \acp{RV} and lower-case counterparts, e.g., $x$, for their realizations. For random vectors, similar notation above is used, e.g., we use $X_a^b$ to denote the random vector $(X_a, X_{a+1},\ldots X_{b})$. The transmission is over \ac{biAWGN} channel with BPSK, i.e., $y_1^N=x_1^N+z_1^N$ where $x_1^N\in\{+1,-1\}^N$ and $y_1^N\in\mathbb{R}^N$ are transmitted and received vectors and $z_1^N$ is an AWGN term whose entries are \ac{i.i.d.} as $\mathcal{N}{(0,\sigma^2)}$.

\subsection{Product Codes}\label{sec:prodcodes}
\subsubsection{Code Construction}
An $m$-dimensional  $(N,k)$ product code $\code$ consists of all $m$-dimensional binary hypercubes which satisfy a linear code constraint along each axis \cite{Elias:errorfreecoding54}. The parameters of the resulting product code are \cite[Sec. 18.2]{macwilliams}
\begin{equation}
	N=\prod_{\ell=1}^m N_\ell, \quad k=\prod_{\ell=1}^m k_\ell, \quad \text{and} \quad R=\prod_{\ell=1}^m R_\ell. \label{eq:parameters}
\end{equation}
where $N_\ell$, $k_\ell$, and $ R_\ell $ are the blocklength, dimension, and rate of the $\ell$-th component code, respectively. Similarly, a generator matrix of the resulting product code is \cite[Sec. 18.2]{macwilliams}
\begin{equation}
	\G = \boldsymbol{G}_{1} \otimes \boldsymbol{G}_{2} \otimes \ldots \otimes \boldsymbol{G}_{m}\label{eq:generator_product_base}
\end{equation}
where $\boldsymbol{G}_{\ell}$ is the generator matrix of the $\ell$-th component code. 

\subsubsection{Encoding}\label{sec:prod_enc} Let $v_1^k$ be the $k$-bit message to be encoded. Then, encoding may be performed via the generator matrix obtained in \eqref{eq:generator_product_base} as $c_1^N=v_1^k\G$. Alternatively, product codes can be seen as serially concatenated codes as follows. In the case of $2$-dimensional product codes, which are the topic of interest in this work,\footnote{For algorithms, $2$-dimensional product codes are considered for simplicity. Nevertheless, extensions to $m$-dimensional product codes are trivial.} the information vector $v_1^k$ is rearranged in a $2$-dimensional array by placing its elements row-by-row starting from $v_1$ till $v_k$, where the length of dimension $i\in[2]$ is $k_i$. Then, the subvectors in the $\ell$-th dimension are encoded via the linear component code $\code_\ell$ with parameters $(n_\ell,k_\ell)$. By juxtaposing the rows of the resulting $2$-dimensional array head-to-tail, the codeword $c_1^N$ is obtained.

\subsubsection{Distance Spectrum}
Although the characterization of the complete distance spectrum of a product code is still an elusive problem (with a few exceptions, see, e.g., \cite{A:wef_prod,caire,T02:wef_prod,KG06,CLG+23}), the minimum distance $d$ and the multiplicity of  codewords with weight \MCC{$d$, namely} $A_d$, can be obtained as
\[
d = \prod_{\ell=1}^{m} d_\ell \qquad\text{and}\qquad A_d=\prod_{\ell=1}^m A^{(\ell)}_{d_\ell} 
\]
where $d_\ell$ and $A^{(\ell)}_w$ are the minimum Hamming distance and the multiplicity of the codewords having weight of $w$ in the $\ell$-th component code, respectively. To achieve relatively large minimum distances, product codes based on algebraic component codes are preferred, e.g., \ac{SPC}\cite{rankin1,rankin2,CLG+23}, eBCH\cite{Pyndiah98,Condo22,SLL+23,Yuan23}, \ac{eHamming}\cite{Elias:errorfreecoding54,El-Khamy05,Chiaraluce} and \ac{RM}\cite{SA05,CJL19} codes to name a few. Although the resulting product codes have relatively large minimum distance, they suffer, in general, from large minimum weight multiplicity \cite{Liva2010:Product_Chinacom}.

\subsubsection{Iterative (Turbo) Decoding}
We consider iterative decoding of $2$-dimensional product codes, where \ac{SISO} decoders are assumed to be available for component codes. This is, in practice, achieved either via \ac{BCJR} algorithm or approximated via Chase decoding\cite{chase} modified to generate bit-wise soft output \cite{Pyndiah98}. The decoder iterates between row and column decoding, where the order can be chosen at will. Each component row/column decoding is counted as half iteration. Decoding is terminated either a valid codeword is reached or a maximum number of half iterations $2I_{\mathrm{max}}$ is reached.\footnote{The maximum number of decoding iterations is denoted by $I_{\mathrm{max}}$.}
 
\iffalse
\renewcommand{\arraystretch}{1.3}
\begin{table}
	\caption{Minimum distances and multiplicities of some 
		product codes (\MCC{eH = extended Hamming code}, SPC = single parity check
		code).}
	\begin{center}
	\begin{tabular}{cccc}
		\hline\hline
		$(n,k)$ & $\mathcal{C}_1$ & $\mathcal{C}_2$ & $A_d$ \\
		\hline
		$(128,105,4)$ & SPC $(16,15)$ & SPC $(8,7)$ & $3360$  \\
		$(128,77,8)$ & eH $(16,11)$ & SPC $(8,7)$ & $3920$  \\
		$(256,121,16)$ & eH $(16,11)$ & eH $(16,11)$ & $19600$  \\
		$(256,165,8)$ & eH $(16,11)$ & SPC $(16,15)$ & $16800$  \\
		$(256,225,4)$ & SPC $(16,15)$ & SPC $(16,15)$ & $14400$ \\
		%$(1024,676)$ & eH $(32,26)$ & eH $(32,26)$ & $16$ & $1537600$  \\
		%$(1024,806)$ & eH $(32,26)$ & SPC $(32,31)$ & $8$ & $615040$ \\
		%$(1024,961)$ & SPC $(32,31)$ & SPC $(32,31)$ & $4$ & $246016$ \\
		\hline\hline
	\end{tabular}
\end{center}
	\label{tab:min_dist}
\end{table}
\renewcommand{\arraystretch}{1}
\fi
\subsection{Precoded Polar Codes}\label{sec:polar}

\subsubsection{Code Construction}
The construction of an $(N,k)$ precoded polar code, where $N = 2^n$, starts by defining
\[
\boldsymbol{\mathsf{G}}_{N}=\boldsymbol{\mathsf{B}}_N\GK_2^{\otimes n}
\]
where $\boldsymbol{\mathsf{B}}_N$ denotes the $N\times N$ \emph{bit reversal} matrix \cite{arikan2009channel} and the Kronecker power $\GK_{2}^{\otimes n}$ denotes the $n$-fold Kronecker product of the Hadamard kernel
\[
\GK_{2}\triangleq\begin{bmatrix}
	1 & 0 \\ 
	1 & 1
\end{bmatrix}.
\]
with $\GK_2^{\otimes 1}\triangleq \GK_2$. Then, a generator matrix $\G$ of a precoded polar code is obtained by multiplying $\boldsymbol{\mathsf{G}}_{N}$ from left by $k\times N$ precoding matrix $\boldsymbol{P}$ with full rank, i.e., we have
\[
\G=\boldsymbol{P}\boldsymbol{\mathsf{G}}_{N}.
\]

We restrict our attention to precoding matrices such that
\begin{enumerate}
	\item the first $1$ appearing in each row is the only non-zero element in the column contains that $1$ and
	\item  the column index of first $1$ in a row $i$ cannot be larger than that of first $1$ in row $j$ if $j>i$.
\end{enumerate}
For such precoding matrices, which are called SC-aimed\cite{MV20}, it is immediate to infer the information positions, frozen bits as well as the dynamic frozen bits with the corresponding constraints uniquely. More precisely, the set containing the column indices of the first $1$s in each row is the information set $\mathcal{A}$ and the frozen positions are contained in set $\mathcal{F}\triangleq [N]\setminus \mathcal{A}$. The elements corresponding to the indices of the columns with non-zero elements form a subset $\mathcal{F}_d$ of $\mathcal{F}$, containing the dynamic frozen bit indices and the positions of non-zero elements in those columns define the constraints, i.e., linear functions $f_i\left(u_1^{i-1}\right)$, $\forall i\in\mathcal{F}_d$. Note that the standard frozen bit constraint is a special case where $f_i=0$, $\forall i\in\mathcal{F}\setminus\mathcal{F}_d$.

\subsubsection{Encoding} Encoding is performed in two stages. First, the vector $v_1^k$ of length-$k$ containing the message is multiplied by precoding matrix $\boldsymbol{P}$ from right, resulting in the $N$-bit vector $u_1^N$. Polar transform is applied to $u_1^N$, which outputs the codeword $c_1^N$ to be transmitted upon modulation.

\subsubsection{\ac{SC}/\ac{SCL} Decoding} Given channel output $y_1^N$, \ac{SC} decoding estimates bits successively from $i=1$ to $i=N$ as
\begin{equation}
	\hat{u}_i = \left\{\begin{array}{lll}
		f_i(\hat{u}_1^{i-1}) &\text{if} \,\,\, i\in\mathcal{F} \\
		d_i(y_1^N,\hat{u}_1^{i-1})& \text{otherwise,}
	\end{array}
	\right.
	\label{eq:decision}
\end{equation}
where the decision functions are defined as
\begin{equation}
	d_i(y_1^N,\hat{u}_1^{i-1}) \triangleq \left\{\begin{array}{ll}
		0& \text{if}\,\,\, P_{U_i|Y_1^N,U_1^{i-1}}(0|y_1^N,\hat{u}_1^{i-1})\geq \frac{1}{2}\\
		1& \text{otherwise.}
	\end{array}
	\right.
	\label{eq:decision_fnc2}
\end{equation} 
%\frac{P_{U_i|Y_1^N,U_1^{i-1}}(0|y_1^N,\hat{u}_1^{i-1})}{P_{U_i|Y_1^N,U_1^{i-1}}(1|y_1^N,\hat{u}_1^{i-1}) }\geq1
The myopic probabilities $P(x|y_1^N,\hat{u}_1^{i-1})$, $x\in\{0,1\}$,\footnote{We omit the subscript whenever it is clear from the context.} are computed recursively via \ac{SC} decoding under the assumption that $U_j$, $i<j\leq N$, are \ac{i.i.d.} uniform random bits, where the recursion continues down to initial probabilities obtained from the channel output as
\begin{equation}\label{eq:initial_probs}
	P(x|y_i) \triangleq \frac{p(y_i|x)}{\sum_{x'}p(y_i|x')}, \, x\in\{+1,-1\},\, \forall i\in[N].
\end{equation}
A block error occurs if and only if $\hat{u}_{\mathcal{A}}\neq u_{\mathcal{A}}$, where $u_{\mathcal{A}}$ denotes the subvector $(u_i:i\in\mathcal{A})$. Hence, \ac{SC} decoding is a complete decoder, i.e., it always outputs a valid codeword.

For any partial input sequences $\tilde{u}_{1}^{i}\in\left\{ 0,1\right\}^{i}$, $i\in[N]$, one can  recursively compute quantities
\begin{equation}\label{eq:myopic_prob}
	P^{(i)}(\tilde{u}_{1}^{i}|y_1^N) \triangleq  P^{(i-1)}(\tilde{u}_{1}^{i-1}|y_{1}^{n}) P(\tilde{u}_{i}|y_{1}^{n},\tilde{u}_{1}^{i-1})
\end{equation}
where the right-most term can be computed efficiently via SC decoding operations with the initial value $P^{(0)}(\tilde{u}_{1}^{0}|y_1^N)\triangleq 1$. The suboptimality of \ac{SC} decoding is overcome in practice by \ac{SCL} decoding\cite{tal15} which computes the quantities \eqref{eq:myopic_prob} for several partial input sequences which obey the constraints imposed by frozen bits, also called decoding paths. The set of decoding paths after $i$-th decoding stage is denoted as $\mathcal{L}_i$ and its size is limited by a parameter called list size, denoted by $L$. After the $N$-th decoding stage, the list $\mathcal{L}_N$ contains $L$ paths $\tilde{u}_{1}^{N}$, where each path corresponds to a valid codeword, with associated probabilities $P^{(N)}(\tilde{u}_{1}^{N}|y_1^N)$. As each $\tilde{u}_{1}^{N}$ correspond to a valid codeword $\tilde{c}_{1}^{N}$, we write $P(\tilde{c}_{1}^{N}|y_1^N)= P^{(N)}(\tilde{u}_{1}^{N}|y_1^N)$. Finally, the estimate $\hat{c}_{1}^{N}$ is chosen as the one with the largest probability $P(\tilde{c}_{1}^{N}|y_1^N)$ among all members of the final list $\mathcal{L}\triangleq\{c_1^N:c_1^N=u_1^N\boldsymbol{P}\boldsymbol{\mathsf{G}}_{N},\forall u_1^N\in\mathcal{L}_N\}$.
the one with the largest probability $P(\tilde{c}_{1}^{N}|y_1^N)$ among all members of the final list $\mathcal{L}\triangleq\{c_1^N:c_1^N=u_1^N\boldsymbol{P}\boldsymbol{\mathsf{G}}_{N},\forall u_1^N\in\mathcal{L}_N\}$.
	\section{Non-Systematic Precoded Polar Product Codes}\label{sec:dpolar_product}
Although being one of the best performing codes for the short blocklength regime, the precoded polar codes have not been yet considered to construct product codes. This might be mainly due to the fact that they do not attain an efficient \ac{SISO} decoding, which is of paramount importance for obtaining good performance via product codes. In the following, this gap is filled by proposing precoded polar product codes, which can match the distance properties (restricted to minimum distance terms) of the product codes with \ac{eBCH} component codes, are proposed. The particular selection enables the use of the final list in SCL decoding to obtain \ac{SISO} component code decoding. Decoding latency can be improved significantly compared to decoding an entire precoded polar code similar to the results presented in \cite{CBHL20} for the case of polar codes. As it will be illustrated in Section \ref{sec:numerical_results}, a significant outcome is that the proposed product codes outperform 5G \ac{LDPC} codes at low rates, especially when the error probabilities are low. 
\subsection{Code Construction}
Consider a $2$-dimensional product code where the component code in the $\ell$-th dimension is an $(N_\ell=2^{n_\ell},k_\ell)$ precoded polar code $\mathcal{C}_\ell$ with generator matrix $\G_\ell=\boldsymbol{P_\ell}\boldsymbol{\mathsf{G}}_{N_\ell}$, which is in non-systematic form. Using mixed-product property and examining the structure of precoding matrices, a precoded polar code representation of the product code is obtained.
%where $\boldsymbol{P_\ell}$, $\ell\in[m]$, is \ac{SC}-aimed, i.e., it satisfies the properties 1) and 2) in Section \ref{sec:polar} for all $\ell\in[m]$.
\begin{prop}
	A generator matrix of the precoded polar product code, in a non-systematic form, is given by
	\begin{equation}
		\G = \boldsymbol{P}\GK_2^{n_1+n_2+\ldots+n_m}
	\end{equation}
	where $\boldsymbol{P}=\boldsymbol{P}_1\otimes \boldsymbol{P}_2\otimes\ldots\otimes\boldsymbol{P}_m$. If precoding matrices $\boldsymbol{P_\ell}$, $\forall\ell\in[m]$, are \ac{SC}-aimed, i.e., they satisfy the properties 1) and 2) in Section \ref{sec:polar}, then $\boldsymbol{P}$ is SC-aimed as well.
\end{prop}
Sets $\mathcal{A}$, $\mathcal{F}$, $\mathcal{F}_d$ as well as the constraints $f_i$, $\forall i\in\mathcal{F}_d$, of the resulting product code are uniquely available thanks to the structure of the resulting precoding matrix $\boldsymbol{P}$. This potentially enables the use of encoding and decoding algorithms for such product codes with no modification. Nevertheless, such decoding algorithms, in particular, may not be efficient due to the sub-optimal placement of the information and frozen positions, resulting in relatively large list sizes for near-\ac{ML} decoding performance (see, e.g., \cite{CJL19} for a similar observation). A more exciting direction is the reverse, i.e., to represent precoded polar codes as (irregular) product codes with precoded polar component codes as provided for polar codes in\cite{CBHL20}.
\subsection{Encoding}
For encoding, vector $u_1^N$ is formed by setting information bits as $u_\mathcal{A}=v_1^k$ and frozen bits as $u_i=0$, $\forall i\in\mathcal{F}\setminus\mathcal{F}_d$ and $u_i=f_i\left(u_1^{i-1}\right)$, $\forall i\in\mathcal{F}_d$. Observe that this requires computations only for the bits corresponding to indices in $\mathcal{F}_d$. Then, the codeword $c_1^N$ is obtained by applying polar transform.\footnote{For the considered cases in Section \ref{sec:numerical_results}, cost of precoding is negligible compared to $\frac{1}{2}N\log_2N$ binary operations for polar transform.} Alternatively, the encoding can be performed by placing the message bits into $2$-dimensional array as described in Section \ref{sec:prodcodes} and then encode using row and column precoded polar code encoders consecutively. Both alternatives provide very similar complexity and latency. We assume now that $c_1^N$ is transmitted via BPSK modulation. 

\subsection{Iterative Decoding via \ac{SCL} Component Decoding}\label{sec:iterative_SCL}

In this section, iterative decoding based on turbo principle\cite{Pyndiah98} is proposed, where precoded polar component codes are decoded via modified \ac{SCL} decoding to generate bit-wise soft outputs similar to \cite{CBHL20}. The iterative decoding of \cite{CBHL20} applies row and column decoding simultaneously and the process is terminated early if both row and column decoders output the same codeword. In the following, we take a step back and employ row and column \ac{SCL} decoding (with modifications to obtain soft output) consecutively as in \cite{Pyndiah98} and the process is terminated early if the component code decoding outputs a valid product codeword, which reduces the average number of iterations.\footnote{Note that \cite{CBHL20} provides lower average as well as worst-case latency for the same maximum iterations if there is an access to twice as many computational resources.} In addition, the proposed modifications for the computation of the bit-wise soft-outputs improve the performance and convergence of the iterative decoding.

Formally, upon receiving the channel output $y_1^N$, the proposed decoder reshapes it as $N_1\times N_2$ matrix $\boldsymbol{Y}$ by filling it row-by-row starting from $y_1$ till $y_N$. Then, the \ac{LLR} matrix is obtained via
\begin{equation}
	\boldsymbol{L}^{\mathrm{ch}} = \frac{2}{\sigma^2}\boldsymbol{Y}\label{eq:channel_LLR}
\end{equation}
where the operations are element-wise. In addition, the a-priori \ac{LLR} matrix, which is of the same size as $\boldsymbol{L}^{\mathrm{ch}}$, is set to all-zero matrix, i.e., $\boldsymbol{L}^{\mathrm{a}}=\boldsymbol{0}$, before the first iteration (all coded bits are uniformly distributed). Assume that decoding starts row-wise without the loss of generality. Then, \ac{SCL} decoding is applied to each row with input $\boldsymbol{L}^{\mathrm{in}}\triangleq\boldsymbol{L}^{\mathrm{ch}}+\boldsymbol{L}^{\mathrm{a}}$, which generates $L$ valid codewords per row code stored in $N_2$ lists $\mathcal{L}^{(1)},\ldots,\mathcal{L}^{(N_2)}$, where the superscript in parenthesis correspond to row index. Let $\mathcal{L}^{(i)}_{j,c_j}$ be the subset of $\mathcal{L}^{(i)}$ composed of the codewords where $j$-th position has value $c_j$. Then, an approximation for the soft output of the $j$-th bit of $i$-th row is computed as
\begin{equation}
	L^{\mathrm{app}}_{i,j} = \log\frac{\sum_{c_1^N\in\mathcal{L}^{(i)}_{j,0}} P\left(c_1^N|y_1^N\right)}{\sum_{c_1^N\in\mathcal{L}^{(i)}_{j,1}} P\left(c_1^N|y_1^N\right)}
	\label{eq:soft}
\end{equation}
where $\log$ is the natural logarithm and this computation is valid only if $\mathcal{L}^{(i)}_{j,c_j}$ is non-empty, $\forall c_j\in\{0,1\}$. Otherwise, the soft-output is set by the largest magnitude in the list as
\begin{equation}
	L^{\mathrm{app}}_{i,j} = (1-2c_j) \left|\log\max_{c_1^N\in\mathcal{L}^{(i)}} P\left(c_1^N|y_1^N\right)\right|
	\label{eq:soft_max}
\end{equation}
where $c_j$ is the bit-value at which all the list members agree on and it determines the sign. Suppose that $\hat{\boldsymbol{C}}$ is obtained by taking hard-decision on each element of matrix $\boldsymbol{L}^{\mathrm{app}}$. If the vector $\hat{c}_1^N$ obtained by juxtaposing the rows of $\hat{\boldsymbol{C}}$ is a valid product codeword, which can be easily verified by the use of either $\boldsymbol{P}$ or $\boldsymbol{P}_1$ and $\boldsymbol{P}_2$, then decoding is terminated. Otherwise, the extrinsic \ac{LLR} matrix is computed as
 \begin{equation}
 	\boldsymbol{L}^{\mathrm{e}} = \boldsymbol{L}^{\mathrm{app}}-\boldsymbol{L}^{\mathrm{ch}}-\boldsymbol{L}^{\mathrm{a}}
 	\label{eq:extrinsic}
 \end{equation}
which is used to set the a-priori \ac{LLR} matrix for the second (half) iteration, i.e., $i=2$, as
\begin{equation}
	\boldsymbol{L}^{\mathrm{a}}=\alpha_i\boldsymbol{L}^{\mathrm{e}}\label{eq:a-priori_ith}
\end{equation}
where $\alpha_i$ is a post-processing parameter for the $i$-th iteration as in \cite{Pyndiah98}. Next, the same steps are applied column-wise. For instance, the modified SCL decoding takes the sum of $\boldsymbol{L}^{\mathrm{ch}}$ and $\boldsymbol{L}^{\mathrm{a}}$, given by \eqref{eq:channel_LLR} and \eqref{eq:a-priori_ith}, respectively, as the input, and it is applied column-wise to the precoded polar component codes. Then, it provides soft messages via \eqref{eq:soft}, \eqref{eq:soft_max} and \eqref{eq:extrinsic} as the output with suitable modifications to the indices. If no valid product codeword is obtained and if the maximum number of iterations is not reached, the process continues with row processing, and so on. Finally, decoding is terminated once the total number of iterations is reached to $I_{\mathrm{max}}$.

\subsection{On the Choice of Component Code Precoding}

When using a product code based on, e.g., \ac{eHamming} component codes, it was observed that the block error probability under iterative decoding via \ac{BCJR} or modified Chase component code decoding can be well approximated by \ac{TUB} (in the tighter form)
\begin{equation}
	P_B \simeq \frac{1}{2}A_d \mathrm{erfc}\left(\sqrt{dR\frac{E_b}{N_0}}\right) \label{eq:tub}
\end{equation} 
already at moderate error rates \cite{Chiaraluce}, which suggests simply adopting component codes with as large minimum distance as possible, e.g., \ac{eBCH} codes are highly favorable in this sense. However, among other reasons, unavailability of \ac{SISO} component code decoding hinders the potential for approaching the performance predicted by the \ac{TUB}. \ac{SISO} decoders, e.g., \ac{BCJR}, in general, have exponential complexity in $(N-k)$ for the component code, which brings us to the selection of component codes for SCL decoding.

Eqs. $\eqref{eq:soft}$ and $\eqref{eq:soft_max}$ convert \ac{SCL} decoding into a soft-output algorithm, where its reliability improves with the list size $L$. We aim at choosing short precoded polar component codes with the best possible minimum distance for given $(N_\ell,k_\ell)$, which is also as suited as possible for \ac{SCL} decoding, i.e., it contains majority of the total probability in the final list $\mathcal{L}$ even when decoded with small list sizes.
\begin{example}
	Consider $(16,7)$ \ac{eBCH} code with \ac{WEF}
	\begin{equation}
		A_{\mathrm{eBCH}}(x)=x^{16}+48x^{10}+30x^8+48x^6+1.\label{eq:eBCH_wef}
	\end{equation}
	It is represented as precoded polar code with a suitably chosen precoding matrix $\boldsymbol{P}_{\mathrm{eBCH}}$, which imposes information set $\mathcal{A}_{\mathrm{eBCH}}=\{4,7,8,12,14,15,16\}$ and the dynamic frozen bit constraints $f_6(u_1^5)=u_4$, $f_{10}(u_1^9)=f_{11}(u_1^{10})=u_4\oplus u_7$ and $f_{13}(u_1^{12})=u_7$. By applying the guidelines of Miloslavskaya et. al.\cite{MV20}, the precoding matrix $\boldsymbol{P}_{\mathrm{opt.}}$ with the same \ac{WEF} as given by $A_{\mathrm{eBCH}}(x)$ and a more suited selection of the information positions for \ac{SCL} decoding is found. The optimized precoding matrix imposes $\mathcal{A}_{\mathrm{opt.}}=\{6,7,8,12,14,15,16\}$ and  $f_{10}(u_1^9)=u_6\oplus u_7$ and $f_{11}(u_1^{10})=u_6$. Note that there is no $7\times 16$ precoding matrix which provides as large minimum distance as $6$ with better \ac{SC} decoding performance than $\boldsymbol{P}_{\mathrm{prop}}$.
\end{example}
As motivated by Example 1, we propose to directly use the precoding matrices optimized for SCL decoding by Miloslavskaya et. al.\cite{MV20} to illustrate the potential of precoded polar product codes, providing remarkable performance in combination with iterative decoding described above. In particular, two precoded polar codes from \cite[Table 1]{MV20} with parameters $(32,17)$ and $(32,21)$ are selected, both with minimum distances of $6$ compared to $4$ of polar codes with the same parameters. This means the product code minimum distances improve from $16$ to $36$ for all considered cases, which hints significant performance improvement if proposed iterative decoding performs well. Note finally that the $(32,21)$ precoded polar code has the same distance spectrum as $(32,21)$ eBCH code as illustrated for the $(16,7)$ codes in Example 1. %In Table 1, we list the minimum distance obtained by 

	\section{Numerical Results}
\label{sec:numerical_results}
Simulation results are provided for codes of rates $R<0.5$, over the \ac{biAWGN} channel. The results are provided in terms of \ac{CER} vs. \ac{SNR} in $E_b/N_0$. For product codes with polar-like component codes, \ac{SCL} component code decoding is used to generate bit-wise soft outputs as described via Eqs. \eqref{eq:soft} and \eqref{eq:soft_max}. We set $\alpha_1^{2I_{\mathrm{max}}}=\left(\frac{1}{8},\frac{1}{8},\frac{2}{8},\frac{2}{8},\frac{3}{8},\frac{3}{8},\frac{4}{8},\frac{4}{8},\frac{4}{8},\dots\right)$, which provide improvements compared to fixing all the elements to $0.5$ as given, e.g., in \cite{Liva2010:Product_Chinacom}. For product codes, $I_{\mathrm{max}}$ is set to $20$ while the maximum number of BP iterations is set to $100$ for LDPC codes. The \acp{TUB} are also provided for the product codes.

The performance is compared to that of product codes with \ac{eBCH} and polar component codes as well as to that of 5G LDPC codes. In most cases, proposed codes outperform the competitors. As evidenced by the analyzed \acp{TUB}, the improvement in performance by the inclusion of precoding matrices is significant, where the gains reach more than $2$ dB. For the case of $(256,49)$ codes, the performance of the proposed code is the same as that of the product code with \ac{eBCH} component codes, where the latter employed \ac{BCJR} row/column decoding. For the case of $(1024,441)$ product code with \ac{eBCH} component codes, the Chase-Pyndiah algorithm is applied with $p=5$ and the tuning parameters $\alpha$ and $\beta$ given by \cite{Pyndiah98}.\footnote{I want to thank Andreas Stra{\ss}hofer (TUM) for the simulation.} Note that results of Yuan et. al. demonstrates that SOGRAND, a version of \ac{GRAND}\cite{DLM19:GRAND} providing bit-wise soft output, improves its performance significantly thanks to the improved \ac{SISO} component code decoding\cite[Fig. 6]{Yuan23}. Therefore, the performance of product code with \ac{eBCH} component code might get closer to that of precoded polar product code at the expense of larger component code decoding complexity.
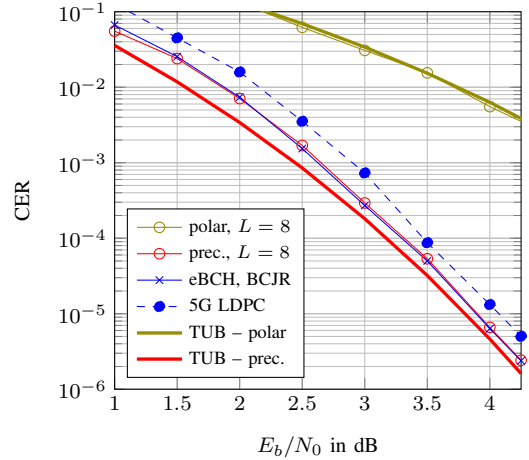
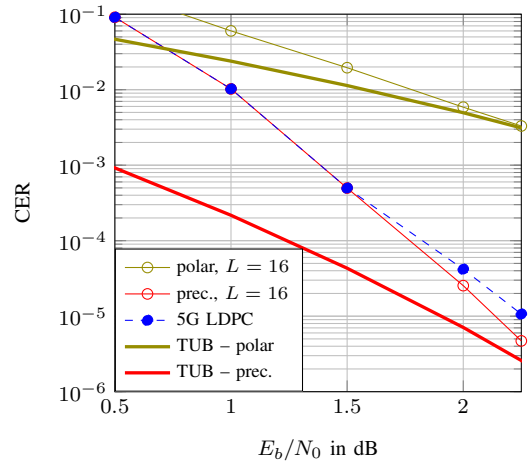
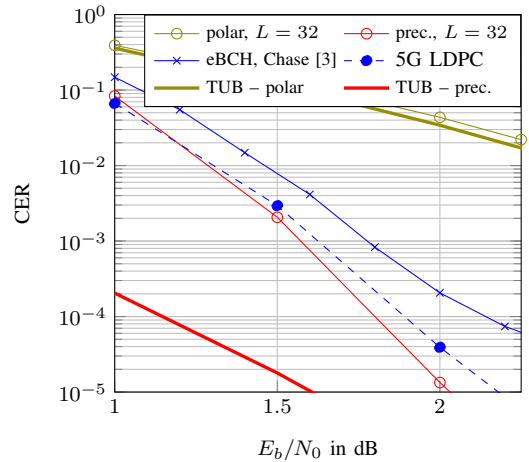
\begin{figure}
	\centering
	\begin{subfigure}{0.49\textwidth}
		\begin{tikzpicture}[scale=1]
	\footnotesize
	\begin{semilogyaxis}[
		legend columns=1,
		legend cell align=left,
		legend pos=south west,
		ymin=0.000001,
		ymax=0.1,
		width=2.75in,
		height=2.6in,
		grid=both,
		xmin = 1,
		xmax = 4.25,
		xlabel = $E_b/N_0$ in dB,
		ylabel = CER,
		xtick={1,1.5,2,2.5,3,3.5,4,4.5}
		]
		
		\addplot[olive, mark = o]
		table[x=snr,y=fer]{snr fer
			2 0.13245
			2.5 0.062112
			3 0.030722
			3.5 0.015504
			4 0.0055353
			4.5 0.0022437
		};\addlegendentry{\scriptsize{polar, $L=8$}}

		\addplot[red, mark = o]
		table[x=snr,y=fer]{snr fer
			1 0.055401662049861
			1.5 0.023969319271333
			2 0.007081149978757
			2.5 0.001691446355779
			3 0.000293163142357
			3.5 0.000053501170071
			4 6.6e-06
			4.25 2.42e-06
		};\addlegendentry{\scriptsize{prec., $L=8$}}
	
	\addplot[blue, mark = x]
	table[x=snr,y=fer]{snr fer
		1 0.066711140760507
		1.5 0.025438819638769
		2 0.007371913011426
		2.5 0.001541164503899
		3 0.000270877915324
		3.5 0.000050022184839
		4 0.000006341415467
		4.25 2.38e-06
	};\addlegendentry{\scriptsize{eBCH, BCJR}}
	
			\addplot[blue, dashed, mark = *]
	table[x=snr,y=fer]{snr fer
		1 0.127713920817369
		1.5 0.044984255510571
		2 0.015931177314004
		2.5 0.003538820864888
		3 0.000732375386328
		3.5 0.0000871534126661798
		4.0 0.000013235668220
		4.25 0.0000050255731314366
	};\addlegendentry{\scriptsize{5G LDPC}}

		\addplot[olive, line width = 1.25pt]
table[x=snr,y=fer]{snr fer
	1 0.395192359649740
	1.5 0.235252627651035
	2 0.132134696988111
	2.5 0.069539400533731
	3 0.034023643317242
	3.5 0.015341014790380
	4 0.006312056446957
	4.5 0.002343833085958
	5 0.000775761694933
	5.5 0.000225695083205
	6 0.000056821551585
};\addlegendentry{\scriptsize{TUB -- polar}}

		\addplot[red, line width = 1.25pt]
table[x=snr,y=fer]{snr fer
	1 0.035823663816487
	1.5 0.011793924021397
	2 0.003410815726695
	2.5 0.000852983374087
	3 0.000181234240348
	3.5 0.000032074108404
	4 0.000004624116695
	4.25 1.6134e-06
};\addlegendentry{\scriptsize{TUB -- prec.}}

%\addplot[blue, dashed, line width = 1.25pt]
%table[x=snr,y=fer]{snr fer
%	1 0.035823663816487
%	1.5 0.011793924021397
%	2 0.003410815726695
%	2.5 0.000852983374087
%	3 0.000181234240348
%	3.5 0.000032074108404
%	4 0.000004624116695
%	4.25 1.6134e-06
%};%\addlegendentry{TUB PPP}
		
	\end{semilogyaxis}
\end{tikzpicture}
		%\vspace{-8mm}
		\caption{$(256, 49)$}
		\label{fig:256_49}
	\end{subfigure}
	%\hfill
	\begin{subfigure}{0.49\textwidth}
		\begin{tikzpicture}[scale=1]
	\footnotesize
	\begin{semilogyaxis}[
		legend columns=1,
		legend cell align=left,
		%legend pos=south west,
		legend style={at={(0,0)},anchor=south west},
		ymin=0.000001,
		ymax=0.1,
		width=2.75in,
		height=2.6in,
		grid=both,
		xmin = 0.5,
		xmax = 2.25,
		xlabel = $E_b/N_0$ in dB,
		ylabel = CER,
		xtick={0.5,1,1.5,2,2.5,3,3.5,4,4.5}
		]

		\addplot[olive, mark = o]
table[x=snr,y=fer]{snr fer
	0.5 0.20202
	1 0.060096
	1.5 0.019558
	2 0.0058685
	2.25 0.0033055
};\addlegendentry{\scriptsize{polar, $L=16$}}
		
		\addplot[red, mark = o]
		table[x=snr,y=fer]{snr fer
			0 0.420168067226891
			0.5 0.091074681238616
			1 0.010234367004401
			1.5 0.000497807159463
			2 0.000025393432596
			2.25 4.706135289500389e-06
		};\addlegendentry{\scriptsize{prec., $L=16$}}
		
		\addplot[blue, dashed, mark = *]
		table[x=snr,y=fer]{snr fer
			0 0.420168067226891
			0.5 0.091074681238616
			1 0.010234367004401
			1.5 0.000497807159463
			2 0.000041945985142
			2.25 1.065930918509475e-05
		};\addlegendentry{\scriptsize{5G LDPC}}
	
			\addplot[olive, line width = 1.25pt]
	table[x=snr,y=fer]{snr fer
		0 0.084928939503435
		0.5 0.046598423511896
		1 0.023889003245862
		1.5 0.011349860842306
		2 0.004951859553877
		2.5 0.001963644801179
		3 0.000699613677434
		3.5 0.000221066935314
		4 0.000061057538828
		4.5 0.000014501435643
	};\addlegendentry{\scriptsize{TUB -- polar}}
	
	\addplot[red, line width = 1.25pt]
	table[x=snr,y=fer]{snr fer
		0 0.003353677197328
		0.5 0.000920512666166
		1 0.000217102981484
		1.5 0.000043194380544
		2 0.000007101224834
		2.5 0.000000942550726
	};\addlegendentry{\scriptsize{TUB -- prec.}}
	\end{semilogyaxis}
\end{tikzpicture}
		%\vspace{-8mm}
		\caption{$(1024, 289)$}
		\label{fig:1024_289}
	\end{subfigure}
	\begin{subfigure}{0.49\textwidth}
		\begin{tikzpicture}[scale=1]
	\footnotesize
	\begin{semilogyaxis}[
		legend columns=2,
		legend cell align=left,
		legend style={at={(1,1)},anchor=north east},
		%legend pos=north east,
		ymin=0.00001,
		ymax=1,
		width=2.75in,
		height=2.6in,
		grid=both,
		xmin = 1,
		xmax = 2.25,
		xlabel = $E_b/N_0$ in dB,
		ylabel = CER,
		xtick={1,1.5,2,2.5,3,3.5,4,4.5}
		]

		\addplot[olive, mark = o]
		table[x=snr,y=fer]{snr fer
			1 0.39216
1.5 0.14065
2 0.043516
2.25 0.022041
		};\addlegendentry{\scriptsize{polar, $L=32$}}
		
		\addplot[red, mark = o]
		table[x=snr,y=fer]{snr fer
			1 0.0834725
			1.5 0.00206407
			2 0.0000133659
			2.25 1.3e-6
		};\addlegendentry{\scriptsize{prec., $L=32$}}
	
		\addplot[blue, mark = x]
	table[x=snr,y=fer]{snr fer
		0.8 3.18e-1
		1 1.48e-1
		1.2 5.5e-2
		1.4 1.48e-2
		1.6 4.14e-3
		1.8 8.3e-4
		2 2.06e-4
		2.2 7.4e-5
		2.4 3.65e-5
	};\addlegendentry{\scriptsize{eBCH, Chase [3]}}
	
			\addplot[blue, dashed, mark = *]
	table[x=snr,y=fer]{snr fer
		1 0.066666666666667
		1.5 0.002944814182225
		2 0.000039170202757
		2.25 0.0000058817
	};\addlegendentry{5G LDPC}
		
				\addplot[olive, line width = 1.25pt]
		table[x=snr,y=fer]{snr fer
			1 0.359231739937550
			1.5 0.118266849214568
			2 0.034202902148248
			2.5 0.008553527723487
			3 0.001817376687937
			3.5 0.000321632031498
			4 0.000046369614637
			4.5 0.000005311685382
		};\addlegendentry{\scriptsize{TUB -- polar}}
		
		\addplot[red, line width = 1.25pt]
		table[x=snr,y=fer]{snr fer
			1 0.000204661010264907
			1.5 0.000017898200786066
			2 0.000001170230049831
			2.5 0.000000055211702895
		};\addlegendentry{\scriptsize{TUB -- prec.}}
		
	\end{semilogyaxis}
\end{tikzpicture}
		%\vspace{-8mm}
		\caption{$(1024, 441)$}
		\label{fig:1024_441}
	\end{subfigure}
\caption{\ac{CER} vs. \ac{SNR} over the \ac{biAWGN} channel for the proposed codes under the described iterative decoding, where $I_{\mathrm{max}}=20$. The performance is compared to those of product codes with \ac{eBCH} component codes (with \ac{BCJR} for (a) and modified Chase for (c) with $20$ iterations), polar product codes and 5G NR LDPC codes (under BP decoding with $100$ iterations).}
\label{fig:comp}
\end{figure}

	\section{Conclusions}
\label{sec:conc}

Under iterative decoding, low-rate precoded polar product codes provide competitive performance to state-of-art ones, e.g., LDPC codes of 5G or product codes with eBCH component codes. Future works should focus on longer blocklengths and improved \ac{SISO} component code decoding for lower complexity, which might provide potential for unified coding/decoding for future standards.

%\newpage
%\bibliography{IEEEabrv,product}

\begin{thebibliography}{10}
\providecommand{\url}[1]{#1}
\csname url@samestyle\endcsname
\providecommand{\newblock}{\relax}
\providecommand{\bibinfo}[2]{#2}
\providecommand{\BIBentrySTDinterwordspacing}{\spaceskip=0pt\relax}
\providecommand{\BIBentryALTinterwordstretchfactor}{4}
\providecommand{\BIBentryALTinterwordspacing}{\spaceskip=\fontdimen2\font plus
\BIBentryALTinterwordstretchfactor\fontdimen3\font minus
  \fontdimen4\font\relax}
\providecommand{\BIBforeignlanguage}[2]{{%
\expandafter\ifx\csname l@#1\endcsname\relax
\typeout{** WARNING: IEEEtran.bst: No hyphenation pattern has been}%
\typeout{** loaded for the language `#1'. Using the pattern for}%
\typeout{** the default language instead.}%
\else
\language=\csname l@#1\endcsname
\fi
#2}}
\providecommand{\BIBdecl}{\relax}
\BIBdecl

\bibitem{Elias:errorfreecoding54}
P.~Elias, ``Error-free coding,'' \emph{IRE Trans. Inf. Theory}, vol. PGIT-4,
  pp. 29--37, Sep. 1954.

\bibitem{Abramson}
N.~{Abramson}, ``Cascade decoding of cyclic product codes,'' \emph{IEEE Trans.
  Commun. Tech.}, vol.~16, no.~3, pp. 398--402, Jun. 1968.

\bibitem{Pyndiah98}
R.~M. Pyndiah, ``Near-optimum decoding of product codes: block turbo codes,''
  \emph{{IEEE} Trans. Commun.}, vol.~46, no.~8, pp. 1003--1010, Aug. 1998.

\bibitem{Berrou93}
C.~Berrou, A.~Glavieux, and P.~Thitimajshima, ``Near {Shannon} limit
  error-correcting coding and decoding: {Turbo}-codes. 1,'' in \emph{Proc. IEEE
  Int. Conf. on Commun.}, vol.~2, 1993, pp. 1064--1070.

\bibitem{NHB97}
H.~Nickl, J.~Hagenauer, and F.~Burkert, ``Approaching shannon's capacity limit
  by 0.27 db using hamming codes in a 'turbo'-decoding scheme,'' in \emph{Proc.
  of IEEE Int. Symp. on Inf. Theory}, 1997, pp. 12--.

\bibitem{Li04}
J.~Li, K.~R. Narayanan, and C.~N. Georghiades, ``Product accumulate codes: a
  class of codes with near-capacity performance and low decoding complexity,''
  \emph{IEEE Trans. Inf. Theory}, vol.~50, no.~1, pp. 31--46, Jan. 2004.

\bibitem{JPN17}
Y.-Y. Jian, H.~D. Pfister, and K.~R. Narayanan, ``Approaching capacity at high
  rates with iterative hard-decision decoding,'' \emph{IEEE Trans. Inf.
  Theory}, vol.~63, no.~9, pp. 5752--5773, 2017.

\bibitem{AM02}
C.~Argon and S.~McLaughlin, ``A parallel decoder for low latency decoding of
  turbo product codes,'' \emph{IEEE Commun. Lett.}, vol.~6, no.~2, pp. 70--72,
  2002.

\bibitem{Chiaraluce}
F.~Chiaraluce and R.~Garello, ``Extended {H}amming product codes analytical
  performance evaluation for low error rate applications,'' \emph{{IEEE} Trans.
  Wireless Commun.}, vol.~3, no.~6, pp. 2353--2361, Nov. 2004.

\bibitem{BCJR}
L.~Bahl, J.~Cocke, F.~Jelinek, and J.~Raviv, ``Optimal decoding of linear codes
  for minimizing symbol error rate,'' \emph{{IEEE} Trans. Inf. Theory},
  vol.~20, no.~2, pp. 284--287, Mar. 1974.

\bibitem{KoetterPC}
X.~{Tang} and R.~{Koetter}, ``Performance of iterative algebraic decoding of
  codes defined on graphs: An initial investigation,'' in \emph{IEEE Inf.
  Theory Workshop}, Sep. 2007, pp. 254--259.

\bibitem{Haeger2018}
C.~{H\"ager} and H.~D. {Pfister}, ``Approaching miscorrection-free performance
  of product codes with anchor decoding,'' \emph{{IEEE} Trans. Commun.},
  vol.~66, no.~7, pp. 2797--2808, Jul. 2018.

\bibitem{MAS16:TPC}
H.~Mukhtar, A.~Al-Dweik, and A.~Shami, ``Turbo product codes: Applications,
  challenges, and future directions,'' \emph{IEEE Communications Surveys \&
  Tutorials}, vol.~18, no.~4, pp. 3052--3069, 2016.

\bibitem{arikan2009channel}
E.~Ar{\i}kan, ``Channel polarization: A method for constructing
  capacity-achieving codes for symmetric binary-input memoryless channels,''
  \emph{IEEE Trans. Inf. Theory}, vol.~55, no.~7, pp. 3051--3073, Jul. 2009.

\bibitem{tal15}
I.~Tal and A.~Vardy, ``List decoding of polar codes,'' \emph{{IEEE} Trans. Inf.
  Theory}, vol.~61, no.~5, pp. 2213--2226, May 2015.

\bibitem{TM16}
P.~Trifonov and V.~Miloslavskaya, ``Polar subcodes,'' \emph{IEEE Journal on
  Selected Areas in Communications}, vol.~34, no.~2, pp. 254--266, 2016.

\bibitem{arikan2019sequential}
\BIBentryALTinterwordspacing
E.~Ar{\i}kan, ``From sequential decoding to channel polarization and back
  again,'' \emph{CoRR}, vol. abs/1908.09594, 2019. [Online]. Available:
  \url{http://arxiv.org/abs/1908.09594}
\BIBentrySTDinterwordspacing

\bibitem{Coskun18:Survey}
M.~C. Co\c{s}kun, G.~Durisi, T.~Jerkovits, G.~Liva, W.~Ryan, B.~Stein, and
  F.~Steiner, ``Efficient error-correcting codes in the short blocklength
  regime,'' \emph{Elsevier Phys. Commun.}, vol.~34, pp. 66--79, Jun. 2019.

\bibitem{MV20}
V.~Miloslavskaya and B.~Vucetic, ``Design of short polar codes for {SCL}
  decoding,'' \emph{IEEE Trans. Commun.}, vol.~68, no.~11, pp. 6657--6668,
  2020.

\bibitem{KCW+2018}
T.~Koike-Akino, C.~Cao, Y.~Wang, K.~Kojima, D.~S. Millar, and K.~Parsons,
  ``Irregular polar turbo product coding for high-throughput optical
  interface,'' in \emph{2018 Optical Fiber Communications Conference and
  Exposition (OFC)}, 2018, pp. 1--3.

\bibitem{BCL19}
V.~Bioglio, C.~Condo, and I.~Land, ``Construction and decoding of product codes
  with non-systematic polar codes,'' in \emph{2019 IEEE Wireless Communications
  and Networking Conference (WCNC)}, 2019, pp. 1--6.

\bibitem{macwilliams}
F.~J. Macwilliams and N.~J.~A. Sloane, \emph{The Theory of Error-Correcting
  Codes}, 1st~ed.\hskip 1em plus 0.5em minus 0.4em\relax
  Amsterdam:North-Holland, 1978, vol.~16.

\bibitem{A:wef_prod}
R.~Andrew, ``The weight distributions of some product codes,'' in \emph{Proc.
  of IEEE Int. Symp. on Inf. Theory}, 2000, pp. 226--.

\bibitem{caire}
G.~Caire, G.~Taricco, and G.~Battail, ``Weight distribution and performance of
  the iterated product of single-parity-check codes,'' in \emph{IEEE GLOBECOM},
  Nov. 1994, pp. 206--211.

\bibitem{T02:wef_prod}
L.~Tolhuizen, ``More results on the weight enumerator of product codes,''
  \emph{IEEE Trans. Inf. Theory}, vol.~48, no.~9, pp. 2573--2577, 2002.

\bibitem{KG06}
\BIBentryALTinterwordspacing
M.~El{-}Khamy and R.~Garello, ``On the weight enumerator and the maximum
  likelihood performance of linear product codes,'' \emph{CoRR}, vol.
  abscs/0601095, 2006. [Online]. Available:
  \url{http://arxiv.org/abs/cs/0601095}
\BIBentrySTDinterwordspacing

\bibitem{CLG+23}
M.~C. Co\c{s}kun, G.~Liva, A.~{Graell i Amat}, M.~Lentmaier, and H.~D. Pfister,
  ``Successive cancellation decoding of single parity-check product codes:
  Analysis and improved decoding,'' \emph{IEEE Trans. Inf. Theory}, vol.~69,
  no.~2, pp. 823--841, 2023.

\bibitem{rankin1}
D.~M. Rankin and T.~A. Gulliver, ``Single parity check product codes,''
  \emph{{IEEE} Trans. Commun.}, vol.~49, no.~8, pp. 1354--1362, Aug. 2001.

\bibitem{rankin2}
D.~M. Rankin, T.~A. Gulliver, and D.~P. Taylor, ``Asymptotic performance of
  single parity-check product codes,'' \emph{{IEEE} Trans. Inf. Theory},
  vol.~49, no.~9, pp. 2230--2235, Sep. 2003.

\bibitem{Condo22}
C.~Condo, ``Iterative soft-input soft-output decoding with ordered reliability
  bits grand,'' in \emph{2022 IEEE Globecom Workshops (GC Wkshps)}, 2022, pp.
  510--515.

\bibitem{SLL+23}
A.~Straßhofer, D.~Lentner, G.~Liva, and A.~G.~i. Amat, ``Soft-information
  post-processing for chase-pyndiah decoding based on generalized mutual
  information,'' in \emph{Int. Symp. on Topics in Coding (ISTC)}, 2023, pp.
  1--5.

\bibitem{Yuan23}
\BIBentryALTinterwordspacing
P.~Yuan, M.~Médard, K.~Galligan, and K.~R. Duffy, ``Soft-output (so) grand and
  long, low rate codes to outperform 5 {LDPCs},'' \emph{CoRR}, vol.
  abs/2310.10737, 2023. [Online]. Available:
  \url{http://arxiv.org/abs/2310.10737}
\BIBentrySTDinterwordspacing

\bibitem{El-Khamy05}
M.~{El-Khamy}, ``The average weight enumerator and the maximum likelihood
  performance of product codes,'' in \emph{{Int. Conf. on Wireless Networks,
  Commun. and Mobile Computing}}, vol.~2, June 2005, pp. 1587--1592.

\bibitem{SA05}
A.~J. {Salomon} and O.~{Amrani}, ``{Augmented product codes and lattices:
  Reed-Muller codes and Barnes-Wall lattices},'' \emph{{IEEE} Trans. Inf.
  Theory}, vol.~51, no.~11, pp. 3918--3930, Nov. 2005.

\bibitem{CJL19}
M.~C. Co\c{s}kun, T.~Jerkovits, and G.~Liva, ``Successive cancellation list
  decoding of product codes with {Reed-Muller} component codes,'' \emph{IEEE
  Commun. Lett.}, vol.~23, no.~11, pp. 1972--1976, 2019.

\bibitem{Liva2010:Product_Chinacom}
M.~Lentmaier, G.~Liva, E.~Paolini, and G.~Fettweis, ``From product codes to
  structured generalized {LDPC} codes,'' in \emph{Proc. Chinacom}, Beijing,
  China, Aug. 2010.

\bibitem{chase}
D.~Chase, ``Class of algorithms for decoding block codes with channel
  measurement information,'' \emph{IEEE Trans. Inf. Theoryy}, vol.~18, no.~1,
  pp. 170--182, 1972.

\bibitem{CBHL20}
C.~Condo, V.~Bioglio, H.~Hafermann, and I.~Land, ``Practical product code
  construction of polar codes,'' \emph{IEEE Transactions on Signal Processing},
  vol.~68, pp. 2004--2014, 2020.

\bibitem{DLM19:GRAND}
K.~R. Duffy, J.~Li, and M.~Médard, ``Capacity-achieving guessing random
  additive noise decoding,'' \emph{IEEE Trans. Inf. Theory}, vol.~65, no.~7,
  pp. 4023--4040, 2019.

\end{thebibliography}
% Generated by IEEEtran.bst, version: 1.14 (2015/08/26)

\end{document}